\documentclass[doublecol]{epl2}
\usepackage{graphicx}
\usepackage{amsmath}
\usepackage{color}

\title{The dynamical frustration of interlayer excitons delocalizing in bilayer quantum antiferromagnets}
\shorttitle{Dynamical frustration of exciton in bilayer antiferromagnets}

\author{L. Rademaker\inst{1}\thanks{E-mail: \email{rademaker@lorentz.leidenuniv.nl}} \and K. Wu\inst{1,2} \and H. Hilgenkamp\inst{1,3} \and J. Zaanen\inst{1}}
\shortauthor{L. Rademaker \etal}

\institute{
	\inst{1} Institute-Lorentz for Theoretical Physics, Leiden University, PO Box 9506, NL-2300 RA Leiden, The Netherlands \\
	\inst{2} Institute for Advanced Study, Tsinghua University, Beijing, 100084, China \\
	\inst{3} Faculty of Science and Technology and MESA+ Institute for Nanotechnology, University of Twente, P.O. Box 217, NL-7500 AE Enschede, The Netherlands
}


\abstract{
Using the self-consistent Born approximation we study the delocalization of interlayer excitons in the bilayer Heisenberg quantum antiferromagnet.
Under realistic conditions we find that the coupling between the exciton motion and the spin system is strongly enhanced as compared to the case
of a single carrier, to a degree that it mimics the confinement physics of carriers in Ising spin systems. We predict that the `ladder spectrum' associated
with this confinement physics should be visible in the c-axis exciton spectra of insulating bilayer cuprates such as YBa$_2$Cu$_3$O$_6$. Our discovery
indicates that finite density systems of such excitons should show very rich physical behavior.}

\begin{document}

\maketitle

\section{Introduction}

\begin{figure}
 \includegraphics[width=8.6cm]{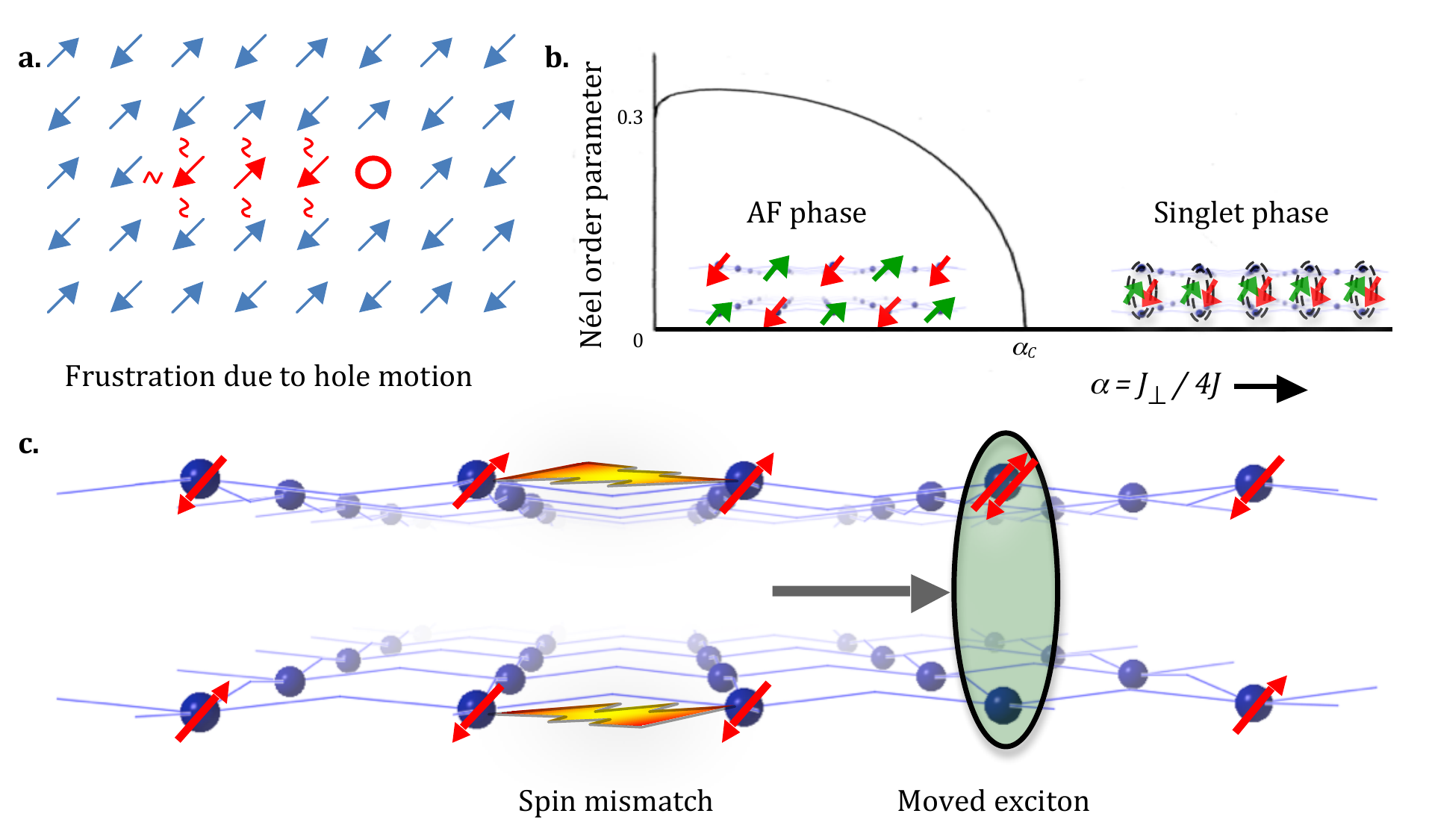}
 \caption{\label{FigBExcitonDynamic}\textbf{a.} A moving hole in a classical single layer Ising antiferromagnet causes a string of spin frustration, which leads to localization.
\textbf{b.} Zero temperature phase diagram of the bilayer Heisenberg model as a function of interlayer coupling strength $\alpha = \frac{J_\perp}{4J}$ on the horizontal axis. At a critical value $\alpha_C$ a quantum phase transition exists from the antiferromagnetic to the singlet phase. The vertical axis shows the N\'{e}el order parameter signaling antiferromagnetism. (Adapted from \cite{Chubukov:1995p2296}.)
\textbf{c.} In a perfect N\'{e}el bilayer, as seen here from the side, the motion of an exciton causes a mismatch in the spin ordering. Here the exciton has moved one position to the right, changing places with the electrons residing on those sites}
\end{figure}

The discovery of  high $T_c$ superconductivity triggered a concerted theoretical effort aimed at understanding the physics of doped Mott
insulators\cite{Imada:1998p2790,Lee:2006p1688}. Although much is still in the dark, the problem of an isolated carrier in the insulator is regarded as well
understood\cite{Bulaevski68,BRINKMAN:1970p5228,SchmittRink:1988p10,Kane:1989p585,Martinez:1991p5159,Dagotto:1994p2661}. It turned out to be a remarkable affair, rooted in the quantum-physical conflict between the antiferromagnetism of the spin system and the delocalizing carrier. This conflict is at its extreme dealing with a classical Ising spin system, where a famous cartoon arises for
the idea of confinement (Fig. \ref{FigBExcitonDynamic}a): the hopping causes a `magnetic string' of overturned spins  between the delocalizing charge and the spin left
at the origin with an exchange  energy increasing linearly in their separation.  It was realized that the quantummechanical
nature of the $S=1/2$ Heisenberg spin system changes this picture drastically. The quantum spin-fluctuations repair efficiently this `confinement damage' in the
spin background and one finds a `spin-liquid polaron' as quasiparticle that propagates coherently through the lattice on a scale set by the exchange
constant. This physics can be reliably addressed by parametrizing the spin system in terms of its linear spin waves (LSW), while the strong coupling
between the spin waves and the propagating hole is well described in terms of the self consistent Born approximation (SCBA). This
turned out to be accurate to a degree that the photoemission results in insulating cuprates were quantitatively explained in this framework\cite{Damascelli:2003p2665}.

A related problem is the delocalization of a bound electron-hole pair (exciton, or more exactly the bound state of a double occupied and vacant site) through the antiferromagnetic background. It is easy to see that
the propagation of an exciton in a single layer is barely affected by the antiferromagnetism since the combined motion of the electron and the hole
neutralize the `damage' in the spin system\cite{Zhang:1998p5176}. A problem of contemporary interest is the exciton formed in a bilayer, where the electron and the hole
reside in the different layers. Using modern growth techniques it appears possible to engineer systems where such excitons are formed at a finite chemical
potential, making it possible to address for instance exciton Bose condensation in the background of an antiferromagnetic spin system.

An exciton in a Mott insulator is, unlike its counterpart in semiconductors, the bound state of the double occupied and vacant sites. We specialize to the case where this Coulomb interaction is large compared to the kinetic energy scale. This might be a realistic assumption in for example the cuprates. Due to Coulomb repulsion between electrons the lowest energy state is always the state where the empty and double occupied site are nearest neighbor along the interlayer rung. Here we report the discovery that under these realistic conditions such bilayer excitons couple extremely strongly through their quantum motion to the spin system.

\begin{figure}
 \includegraphics[width=8.6cm]{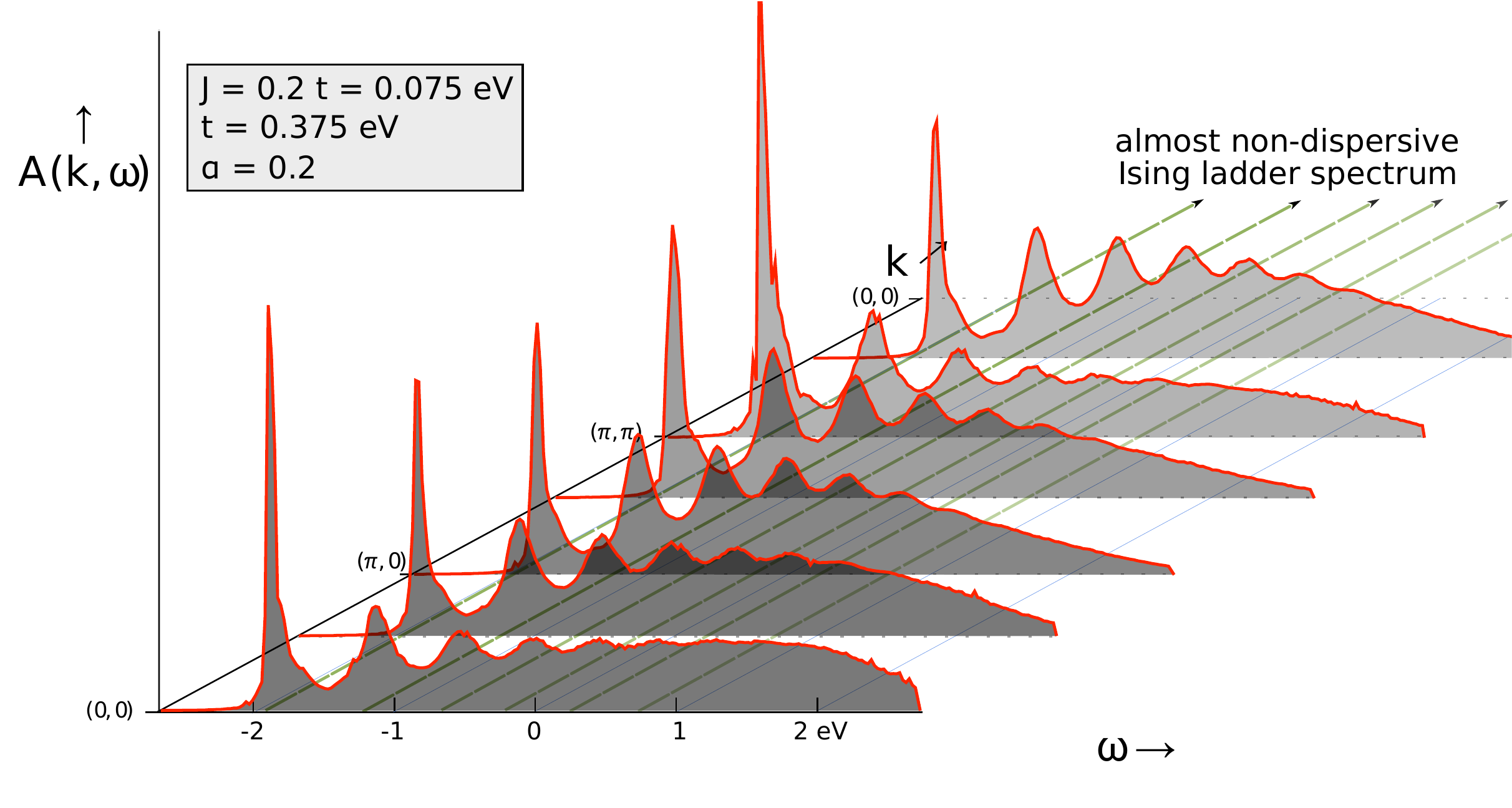}
 \caption{\label{FigIsing}Exciton spectral function for $J=0.2t$ and $\alpha=0.2$. On top of the incoherent bump a strong ladder spectrum has developed, signaling Ising confinement. The exact Ising ladder spectrum is shown in green dotted lines. The Ising peaks are very weakly dispersive, with bandwidth of order $J$.}
\end{figure}

In fact, when the interlayer exchange coupling is small and the exciton hopping rate is large, one enters a regime that is similar to the confinement
associated with the Ising spins, although the spin system is in the quantized Heisenberg regime. This is illustrated by the exciton spectral function
shown in Fig. \ref{FigIsing} as computed with the LSW-SCBA method, showing the non-dispersive  `ladder spectrum'  which is a fingerprint of confinement.
Fig. \ref{FigBExcitonDynamic}c depicts a cartoon of the confinement mechanism: every time the exciton hops it creates two spin flips in the different layers that can only be
repaired by quantum spin fluctuations driven by the interlayer exchange coupling. The rapid intralayer quantum spin fluctuations are now ineffective,
because the restoration of the anti-ferromagnetism requires quantum spin fluctuations that occur simultaneously in the two layers with a probability
that is strongly suppressed.

\begin{figure}
 \includegraphics[width=8.6cm]{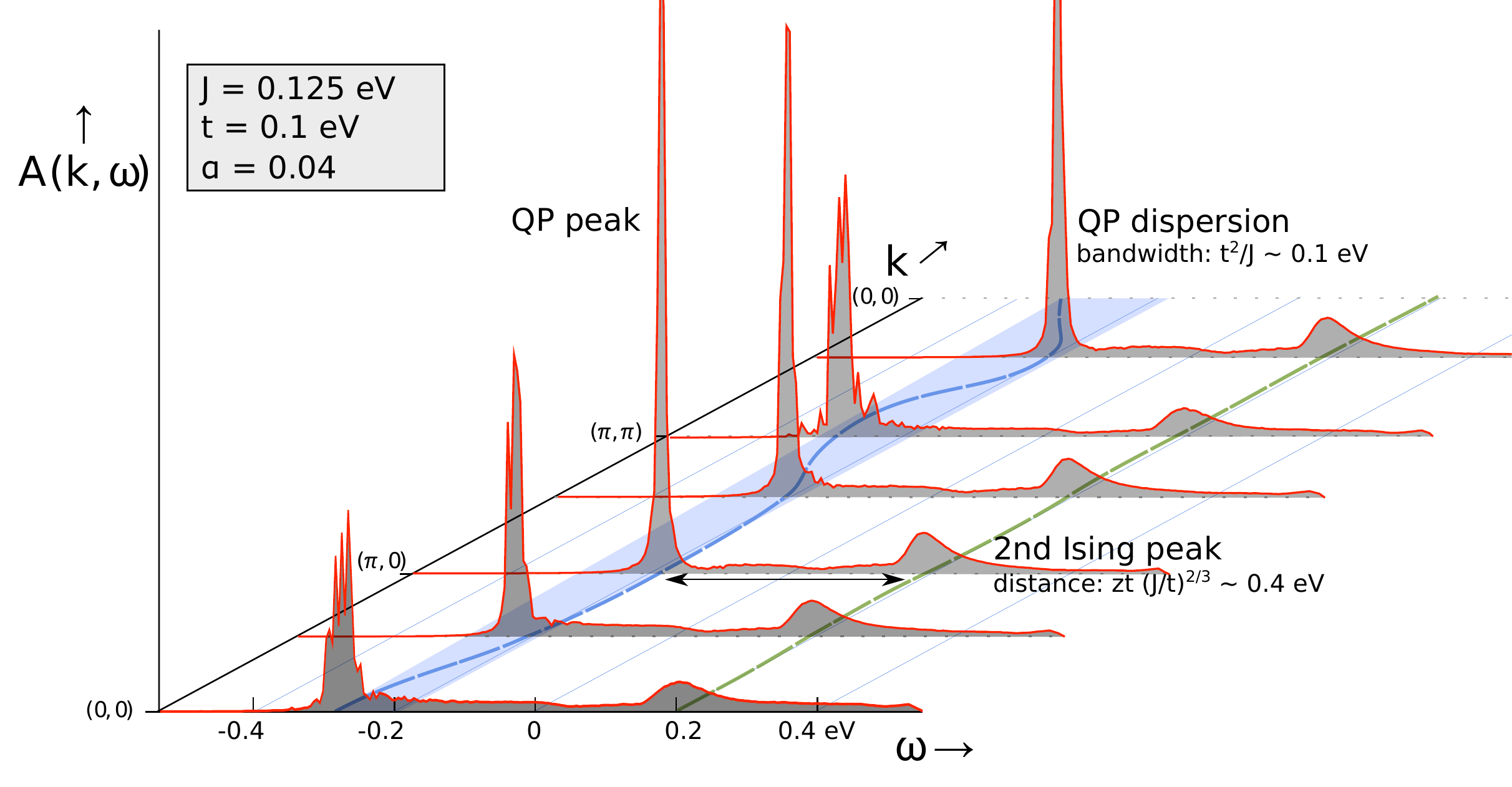}
 \caption{\label{FigYBCO}Expected exciton spectral function for the $c$-axis charge-transfer exciton in YBCO bilayers. We used model parameters $J=0.125$ eV, $t=0.1$ eV and $\alpha=0.04$. The exciton quasiparticle peak has a dispersion with bandwidth $t^2/J$, and the quasiparticle peak is the most pronounced at the line between $(\pi,0)$ and $(0, \pi)$. Following at a distance of $zt (J/t)^{2/3}$, a secondary peak develops as a sign of Ising confinement.}
\end{figure}

This confinement effect can be studied directly in experiment by measuring the exciton spectrum in c-axis optical absorption
of the YBa$_2$Cu$_3$O$_6$ (YBCO) insulating bilayer system. Using realistic parameters we anticipate that this will look like Fig. \ref{FigYBCO}: the main difference with Fig. \ref{FigIsing} is that the exciton hopping rate is now of order of the exchange energy and in this adiabatic regime  the spectral weight in the ladder spectrum
states is reduced. Our discovery is particularly significant in the context of the effort to create bilayer exciton systems at a finite density. Stripes and other
complex ordering phenomena in the finite density electron systems are believed to be driven by the hole-spin  system `quantum frustration'. Since the
excitons are bosons, this holds the promise that the same type phenomena can now be studied in a much more tractable bosonic setting.

\section{The exciton $t-J$ model}

Let us now discuss our calculations.  Our point of departure is the bilayer Heisenberg model \cite{Manousakis:1991p2291},
\begin{equation}
	H_J = J \sum_{<ij> \, l} \vect{S}_{il} \cdot \vect{S}_{jl}
		+ J_\perp \sum_i \vect{S}_{i1} \cdot \vect{S}_{i2}
	\label{BilayerHeisenberg2}
\end{equation}
where $\vect{S}_{il}$ is the spin operator on site $i$ and layer $l = 1,2$.
This model is known to describe the physics of half-filled bilayer Mott insulators, with on each site a localized spin-$\frac{1}{2}$ state. A natural basis for the bilayer spin states consists of singlet and triplet states on each interlayer rung \cite{Chubukov:1995p2296,Duin:1997p2301}. The singlet state $|0 \; 0 \rangle$ equals $\frac{1}{\sqrt{2}} \left( | \uparrow_1 \downarrow_2 \rangle -  | \downarrow_1 \uparrow_2 \rangle \right)$ and similar expressions hold for the triplet states $|1 \; m \rangle$ with $m = -1, 0, +1$.

In the Mott insulator the magnetism is fully determined by the interplay between the interlayer Heisenberg energy $J_\perp$ and the intralayer energy $J$. For low values of $\alpha \equiv J_\perp / zJ$ ($= J_\perp / 4J$ on a square lattice) the ground state is an antiferromagnet, whilst in the limit of infinite $J_\perp$ the electrons on each interlayer rung tend to form singlets. Consequently, the system undergoes a $O(3)$ universality class quantum phase transition at some finite $J_\perp$ from an antiferromagnetic phase to a `singlet' phase. The zero-temperature phase diagram of the bilayer Heisenberg model \cite{Sandvik:1994p3011,Chubukov:1995p2296,Duin:1997p2301} is shown in Figure \ref{FigBExcitonDynamic}b.

Following the premise that the Coulomb interaction is large compared to the kinetic energy, we can write down a low-energy effective exciton model where the doubly occupied and vacant sites are all bound into an interlayer exciton state $| E \rangle$ consisting of a doublon-holon pair on a rung. This exciton can hop around, a process in which the exciton swaps places with the neighboring spin state described by the Hamiltonian
\begin{equation}
	H_t = - t \sum_{<ij>} | E_j \rangle \left(  | 0 \; 0 \rangle_i \langle 0 \; 0 |_j
			+ \sum_m |1 \; m \rangle_i \langle 1  \; m |_j
			 \right) \langle E_i |
		\label{ExcitonHop}
\end{equation}
The sum runs over all nearest neighbor sites $ij$ and the energy $t$ is the exciton hopping energy. Following the same strong coupling perturbation theory that led to the introduction of the $t-J$ model based on the Hubbard model, we can express the exciton hopping parameter $t = t_e^2 / V$ in terms of the bare electron hopping energy $t_e$ and the exciton binding energy $V$. The total Hamiltonian $H_{t-J}$ of this new `exciton $t-J$ model' consists of both the exciton hopping terms (\ref{ExcitonHop}) and the Heisenberg terms (\ref{BilayerHeisenberg2}).

\section{Computations}

Using the exciton $t-J$ model we can compute the exciton propagator
\begin{equation}
	G (k, \omega) = \langle \psi_0 | E_{k} \frac{1}{\omega - H_{t-J} + i \epsilon} E_{k}^\dagger | \psi_0 \rangle	
\end{equation}
where $| \psi_0 \rangle$ is the ground state of the bilayer Heisenberg model, and $E^\dagger_k$ creates an exciton state with momentum $k$. Next we compute the exciton spectral function $A (k, \omega) = - \frac{1}{\pi} \mathrm{Im} \left[ G (k, \omega) \right]$, which is directly related to the $c$-axis optical absorption \cite{BruusFlensberg}
\begin{equation}
	\textrm{Im} \left[ \epsilon^{-1} (q^z, \omega) \right]
	\sim (t_\perp)^2 A(k=0, \omega)
\end{equation}
where $t_\perp$ is the interlayer electron hopping. We expect that the exciton spectrum should be detectable at an energy $V$ below the electron-hole continuum, where $V$ is the binding energy of the exciton. Note that the exciton spectral function can also be measured using Electron Energy Loss Spectroscopy (EELS)\cite{Schnatterly1979} or Resonant Inelastic X-ray Scattering (RIXS)\cite{Ament:2010p5208} probing the $c$-axis.

As can be seen in Figure \ref{FigBExcitonDynamic}b, a moving exciton causes frustration in the previously perfect N\'{e}el state. In order to deal with this frustration in a manner similar to the methods developed for the single layer $t-J$ model\cite{SchmittRink:1988p10,Kane:1989p585}, we first need to construct the correct linear spin wave theory (LSW) for the bilayer Heisenberg model \cite{Vojta:1999p5138}. Following \cite{Chubukov:1995p2296} we define the sum and difference of the spin operators as $\mathbf{s}=\mathbf{S}_1+\mathbf{S}_2$ and $\mathbf{\tilde{s}}=\mathbf{S}_1-\mathbf{S}_2$ respectively. The Heisenberg Hamiltonian (\ref{BilayerHeisenberg2}) now reads
\begin{equation}
  H_J =
  	\frac{1}{2} J\sum_{\langle ij\rangle}
  	(\mathbf{s}_{i}\cdot\mathbf{s}_{j}
		+\mathbf{\tilde{s}}_i\cdot\mathbf{\tilde{s}}_j)
	+ \frac{1}{4} J_{\perp}
		\sum_{i}(\mathbf{s}_{i}^2 - \mathbf{\tilde{s}}_i^2).
	\label{BHM3}
\end{equation}
Next we introduce a mean-field approximation to the ground state on each rung by
\begin{equation}
	| G \rangle_i = \eta_i\cos \chi | 0 \; 0 \rangle_i - \sin \chi |1 \; 0 \rangle_i,
\end{equation}
that interpolates between the N\'{e}el state ($\chi=\pi/4$) and the singlet state ($\chi=0$), and $\eta_i = (-1)^i$ alternates on the two sublattices. The angle $\chi$ needs to be fixed by minimizing the ground state energy for a given $\alpha$. On each rung there are three excited spin states orthogonal to the ground state $|G\rangle_i$,
\begin{eqnarray}
	e^\dagger_{i} & =  &\left( \eta_{i}\sin \chi | 0 \; 0 \rangle_i
			+ \cos \chi | 1 \; 0 \rangle_i \right) \langle G |_i , \label{SW1} \\
	b^\dagger_{i+} & = & | 1 \; 1 \rangle_i \langle G | _i  ,\label{SW2}\\
	b^\dagger_{i-} & =  &| 1 \; -1 \rangle_i \langle G | _i. \label{SW3}
\end{eqnarray}
and consequently we can express the spin sum and difference operators exactly in terms of the spin excitation operators $e^\dagger$ and $b^\dagger_{\pm}$,
\begin{eqnarray}
 \tilde{s}_{i}^+ &=&
 		\sqrt{2}\eta_i[\cos{\chi}(b_{i+}^\dagger-b_{i-})+\sin\chi (b_{i+}^\dagger e_i-b_{i-}e_i^\dagger)], \nonumber\\
  s_{i}^+ &=&
  		-\sqrt{2}[\sin{\chi}(b_{i+}^\dagger+b_{i-})-\cos{\chi}(b_{i+}^\dagger e_i+b_{i-}e_i^\dagger)],  \nonumber \\
	\tilde{s}_i^z &=&
		\eta_i \left[  \sin{2\chi}(1-2e_{i}^\dagger e_{i}-\sum_{\sigma=\pm}b_{i\sigma}^\dagger b_{i\sigma}) \right. \nonumber \\ && \left. -\cos{2\chi}(e_{i}^\dagger +e_{i}) \right], \nonumber\\
  s^z_i &=&
  		b^\dagger_{i+} b_{i+} - b^\dagger_{i-} b_{i-}.
	\label{mHP}
\end{eqnarray}
We rewrite (\ref{BHM3}) in terms of the spin excitation operators (\ref{SW1}-\ref{SW3}). The linear spin wave approximation now amounts to normal ordering the Hamiltonian and neglecting all interaction terms between the spin excitation operators\cite{Dyson}. After performing a Fourier and Bogolyubov transformation, we find the spin wave dispersions
\begin{align}
	\epsilon^T_{\mathbf{k},\pm}
		&={Jz\over2}\sqrt{(\sin^22\chi+2\alpha\cos^2\chi\mp \gamma_\mathbf{k} \cos2\chi)^2-\gamma_\mathbf{k}^2}\label{TSW}\\
	\epsilon^L_{\mathbf{k},\pm}
		&=Jz\sqrt{C(C\mp \gamma_\mathbf{k} \cos^22\chi)}\label{LSW}
\end{align}
with $\gamma_\mathbf{k}=\frac{1}{2}(\cos{k_x}+\cos{k_y})$ and $C=\sin^22\chi+\alpha\cos2\chi$. These dispersions correspond to the known spin wave spectrum\cite{Sandvik:1994p3011,Chubukov:1995p2296,Duin:1997p2301} consisting of two single layer transversal (acoustic) spin waves (\ref{TSW}) derived from the $b_{\pm}^\dagger$ operators and a gapped longitudinal (optical) spin wave (\ref{LSW}) from $e^\dagger$. In the N\'{e}el phase, the transversal modes are gapless Goldstone modes while in the singlet phase all spin waves are degenerate gapped triplet excitations.

\begin{figure}
 \includegraphics[width=8.6cm]{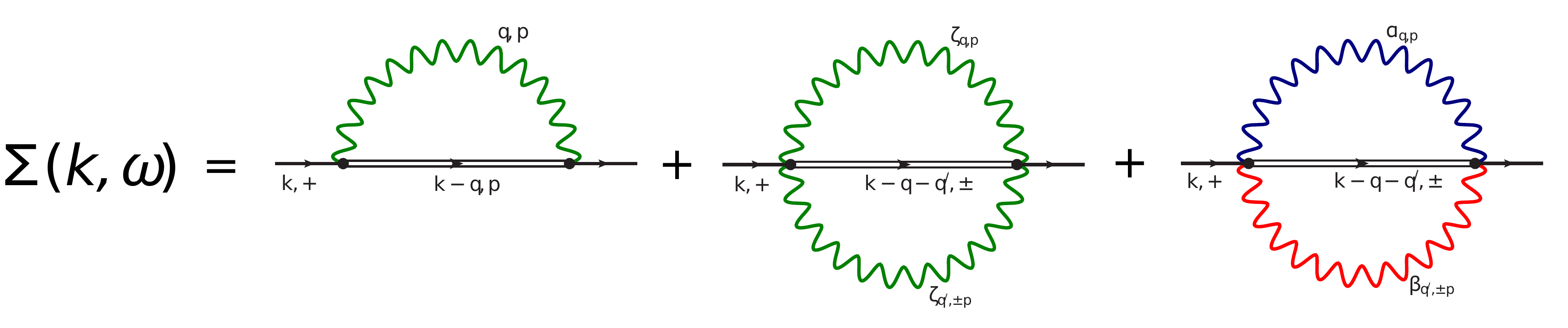}
 \caption{\label{FigSCBA}Feynman diagram representation of the Self-Consistent Born Approximation (SCBA). The self-energy of the exciton depends self-consistently on `rainbow' diagrams where it emits and absorbs either one or two spin waves. The left two diagrams contain interaction with the longitudinal spin wave, the right diagram contains the interaction with the transversal spin waves. Vertex corrections are neglected in the SCBA.}
\end{figure}

Using the expressions for spin excitations (\ref{SW1})-(\ref{SW3}) we find that the exciton hopping term (\ref{ExcitonHop}) takes the form
\begin{eqnarray}
 H_e
 	&=&
		t \sum_{\langle ij\rangle}E_j^\dagger E_i[\cos2\chi(1-e_i^\dagger e_j)\nonumber\\
 && +\sin2\chi(e_i^\dagger+e_j)-\sum_\sigma b_{i\sigma}^\dagger b_{j\sigma}]+h.c. .
\end{eqnarray}
Using the explicit transformations obtained by the Bogolyubov transformation between local spin operators (\ref{SW1})-(\ref{SW3}) and the spin waves we can rewrite the exciton hopping Hamiltonian into an exciton-spin wave interaction term. Subsequently we used the SCBA, which was successfully applied to the single hole problem \cite{SchmittRink:1988p10,Kane:1989p585}, to compute the exciton self-energy. In the SCBA one neglects the vertex corrections of the self-energy, which turned out to be a good approximation in comparison with exact diagonalization studies\cite{Martinez:1991p5159}. Next to the fact that we need to include both types of spin waves, an extra complication arises since the self-energy also contains diagrams with two spin waves. The resulting diagrammatic expression for the exciton self-energy $\Sigma (k, \omega)$ in the SCBA is shown in Figure \ref{FigSCBA}. As for the single layer $t-J$ model, the SCBA can only be solved numerically, and we have obtained the exciton spectral function $A (k, \omega)$ using an iterative procedure with Monte Carlo integration over the spin wave momenta discretized on a 32 $\times$ 32 momentum grid. We start with self-energy $\Sigma=0$ and after approximately 20 iterations the spectral function converged.

\section{Results}

The extensive details of the computations will be published elsewhere\cite{Rademaker2011} and we will illustrate here the resulting exciton spectral function in representative limits. The most important control parameter is $\alpha$ since it is tuning the nature of the spin system, while $t/J$ is just governing the degree of adiabaticity of the exciton motion.  When $\alpha \gg 1$, the magnetic ground state is just a stack of rung singlets which is effectively transparent for the exciton that
propagates freely in this regime  with a bandwidth of order of $8t$. Given the expectation
raised by the cartoon in Fig. \ref{FigBExcitonDynamic}c, the most interesting limit is $\alpha \rightarrow 0$: here the antiferromagnetic order is strongest, while the interlayer
exchange coupling becomes inoperative with regard to flipping back\cite{SchmittRink:1988p10,Kane:1989p585} the two overturned spins
caused by the hop of the exciton. In the strongly adiabatic regime where $t \ll J$ the exciton just localizes since it has to climb a very steep `exchange hill' in order to move at all. Although its spectrum does contain all the `ladder states' associated with its `magnetic string' confinement these
acquire a very small spectral weight because of the adiabatic limit: the first
excited ladder state at energy $Jz$ is just visible with a spectral weight $ \mathcal{O} (t^2/J^2) $.

\begin{figure}
 \includegraphics[width=8.6cm]{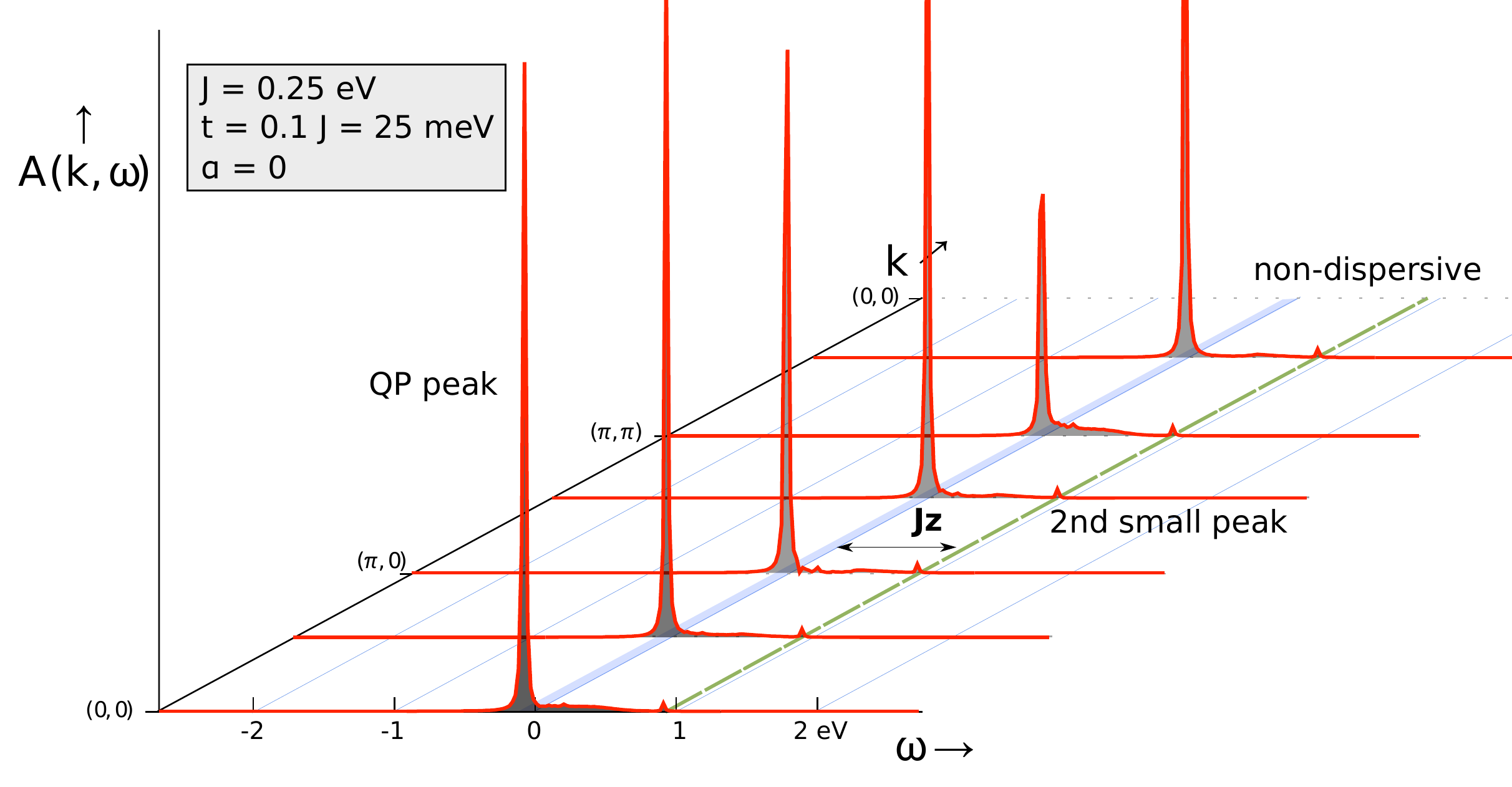}
 \includegraphics[width=8.6cm]{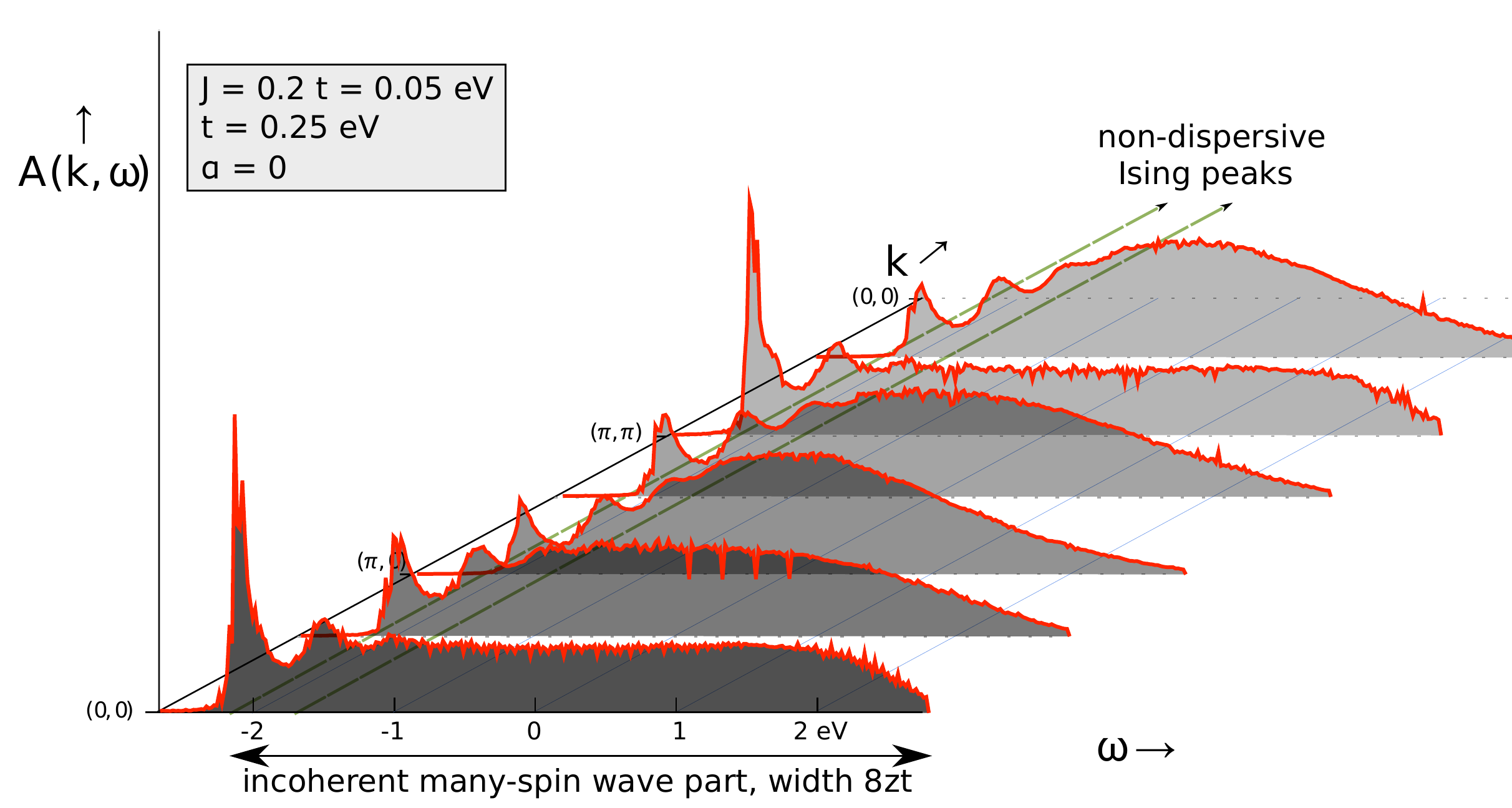}
 \caption{\label{FigAlphaZero}Exciton spectral functions when the interlayer magnetic coupling $\alpha = 0$. In the top figure $(t \ll J)$ the adiabatic motion of the exciton allows for the formation of a clear quasiparticle peak. In the anti-adiabatic limit ($t \gg J$, bottom figure) the spectral weight is shifted into a many-spin wave incoherent bump. Note that both spectra are almost non-dispersive.}
\end{figure}

The confinement effect in this $\alpha \rightarrow 0$ regime becomes fully visible in the anti-adiabatic $t \gg J$ regime. Now the hopping exciton can
explore the confining potential over large distances and the effect is that the spectral weight shifts to the highly excited ladder states leaving
behind a strongly suppressed quasiparticle peak at the bottom of the spectrum. Both the adiabatic and anti-adiabatic results at exactly $\alpha = 0$ are shown in Figure \ref{FigAlphaZero}.

Whereas the ladder spectrum at $\alpha = 0$ seems to be obscured by a broad incoherent bump, inclusion of a finite $\alpha$ seems to intensify the ladder spectrum as is shown in Figure \ref{FigIsing}. To first order in $\alpha$, the interaction of excitons with transversal spin waves is reduced. This favors the formation of a ladder spectrum, which can be seen if one neglects the transversal spin waves in the SCBA. The remaining interaction with the longitudinal spin waves leads up to first order in $\alpha$ to the following self-consistent equation for the exciton self-energy,
\begin{equation}
	\Sigma (\omega) = \frac{\frac{1}{2} z^2 t^2}{\omega - Jz - \Sigma (\omega - Jz)}.
\end{equation}
This is exactly the self-energy associated with a single hole in an Ising magnet\cite{Kane:1989p585}, where the naive picture of the spin frustration as drawn in Figure \ref{FigBExcitonDynamic}c becomes exact. Any motion of the dopant (electron, hole or exciton) away from its reference position increases the energy required for the spin mismatch. Oscillations around the reference position will generate a ladder spectrum, with the energy distance between the two lowest states scaling as $t (J/t)^{2/3}$.

As compared to the Ising spin system the sharp ladder spectrum gets blurred due to the presence of transversal spin waves. The degree to which such blurring occurs depends on the ratio $t/J$. Consequently the ladder spectrum is most visible when $t \ll J$ and at small but finite $\alpha$, as can be seen in Figure \ref{FigIsing}.

The ladder spectrum is, for example, expected for the half-filled cuprate bilayers in YBCO or Bi$_2$Sr$_2$CaCu$_2$O$_{8}$. Following earlier neutron scattering experiments on YBCO\cite{Imada:1998p2790,Tranquada:1989p5209} one can deduce that the effective $J = 125 \pm 5$ meV and $J_\perp = 11 \pm 2$ meV, which corresponds to an effective value of $\alpha = 0.04 \alpha_c$ where $\alpha_c$ is the critical value of $\alpha$\cite{Chubukov:1995p2296}. The question remains what a realistic estimate of the exciton binding energy is. The planar excitons are known to be strongly bound \cite{Zhang:1998p5176} with binding energy of the order of 1-2 eV. Since the Coulomb repulsion scales as $V\sim (\epsilon r)^{-1}$, we can relate the binding energy of the interlayer excitons to that of the planar excitons. The distance between the layers is about twice the in-plane distance between nearest neighbor copper and oxygen atoms, but simultaneously we expect the dielectric constant $\epsilon_c$ along the $c$-axis to be smaller than $\epsilon_{ab}$ due to the anisotropy in the screening. Combining these two effects, we consider it a reasonable assumption that the interlayer exciton binding energy is comparable to the in-plane binding energy. The hopping energy for electrons is approximately $t_e = 0.4$ eV which yields, together with a Coulomb repulsion estimate of $V \sim 1.5$ eV, an effective exciton hopping energy of $t \sim 0.1$ eV. The spectral function corresponding to this parameters is shown in Figure \ref{FigYBCO}. Since $t \sim J$ the ladder spectrum is strongly suppressed compared to the aforementioned anti-adiabatic limit. However, the Ising confinement still shows its signature in a small `second ladder peak' at $0.4$ eV energy above the exciton quasiparticle peak. To the best of our knowledge and to our surprise, the $c$-axis optical conductivity of YBCO has not been measured before. Detection of this second ladder peak in future experiments would suggest that indeed the interlayer excitons in cuprates are frustrated by the spin texture.

\section{Conclusion}

In conclusion, we have shown that the dynamical spin-hole frustration effects that are well known in the context of doped Mott insulators occur
in a strongly amplified form dealing with interlayer excitons in Mott-insulating bilayer systems.  We anticipate that the real significance of this discovery
lies in the potential of creating exciton systems of this kind at a finite density. As compared to the usual doped Mott-insulators, excitons simplify
matters because they are bosons rendering the situation much more tractable. At the same time,  the tendency to exhibit complex ordering
phenomena like the stripes and so forth  are in first instance caused by dynamical frustration effects of the kind discussed in this paper. At least when the magnetic order is maintained,
it appears that the problem of excitons at finite density is sign free and can therefore be addressed
in terms of bosonic field theory. We expect a coexistence of magnetic order and the exciton Bose condensate,
but in the light of the present single exciton results it is clear that these will show unusual strong coupling features.

This gives further impetus to the pursuit to create such finite density correlated exciton systems
in the laboratory. One can wonder whether such physics is already at work in the  four-layer material Ba$_2$Ca$_3$Cu$_4$O$_8$F$_2$ where self-doping effects occur creating simultaneously $p$ and $n$-doped layers\cite{Chen:2006bs}. Much effort has been devoted to create equilibrium finite exciton densities
using conventional semiconductors\cite{Moskalenko:2000p4767}, while exciton condensation has been demonstrated in  coupled semiconductor
2DEGs \cite{Eisenstein:2004go,Butov:2007fr}. In strongly correlated heterostructures, however, formation of finite exciton densities is still far from achieved, although recent developments on oxide interfaces indicate exciting potential (see for example \cite{Pentcheva:2010p5025}). Besides the closely coupled $p$- and $n$-doped conducting interface-layers in these SrTiO$_3$-LaAlO$_3$-SrTiO$_3$ heterostructures, further candidates would be closely coupled $p$- and $n$-doped cuprates, such as YBa$_2$Cu$_3$O$_{7-x}$ or La$_{2-x}$Sr$_x$CuO$_4$ with Nd$_{2-x}$Ce$_x$CuO$_4$. The feasibility of this has already been experimentally demonstrated, e.g. in \cite{Takeuchi95}, but the exact interface effects need to be investigated in more detail, both experimentally as well as theoretically \cite{Ribeiro2006,Millis:2010p5231}.

\acknowledgements
This research was supported by the Dutch NWO foundation through a VICI grant.
The authors thank Jeroen van der Brink, Sergei Mukhin and Aron Beekman for helpful discussions.

\end{document}